\documentclass[iop,apj]{emulateapj}
\slugcomment{{\sc \underline{Accepted for publication in J. of Atmos. and Sol-Terr. Phys. }} Nov 29, 2017}
\usepackage{amsmath,amssymb,amstext}
\usepackage[breaklinks,colorlinks,citecolor=blue,linkcolor=magenta]{hyperref} 

\usepackage[all]{hypcap} 

\usepackage{natbib,color,mhchem}
\shorttitle{Critical frequencies and solar cycle}
\shortauthors{Y\.{I}\u{g}\.{I}t et al.}

\graphicspath{{/Users/erdal/OneDrive/SunfoF_Ali/sunf0f_v07_ali/}
  {./Figures/}}

\begin{document}
\title{\textbf{
Critical frequencies of the ionospheric $F_1$ and $F_2$ layers during the last four
  solar cycles: sunspot group type dependencies}}

\author{Erdal Y\.{I}\u{g}\.{I}t\altaffilmark{1}, Ali Kilcik\altaffilmark{2}, 
Ana Georgina Elias\altaffilmark{3},
Bur\c cin D\"onmez\altaffilmark{2},
Atilla Ozguc\altaffilmark{4},
Vasyl Yurchshyn\altaffilmark{5,6},
Jean-Pierre Rozelot\altaffilmark{7},
}



\altaffiltext{1}{Department of Physics and Astronomy, Space Weather Laboratory, George Mason
  University, Fairfax, Virginia, USA.}

\altaffiltext{2}{Akdeniz University, Department of Space Sciences and Technologies, Turkey}

\altaffiltext{3}{Univ. Nacl. Tucuman, Fac. Ciencias Exactas and Tecnol., Dept. Fis., Argentina}
\altaffiltext{4}{Kandilli Observatory and Earthquake Research Institute, Bogazici University, Turkey}
\altaffiltext{5}{Big Bear Solar Observatory, Big Bear City, CA, USA} 

\altaffiltext{6}{Korea Astronomy and Space Science Institute, Daejeon, Korea}

\altaffiltext{7}{Universit\'e de la Cote d'Azur, Grasse, France.}

\begin{abstract} 
 The long term solar activity dependencies of ionospheric F$_1$ and F$_2$ regions' critical
  frequencies ($f_0F_1$ and $f_0F_2$) are analyzed for the last four solar cycles
  (1976--2015). We show that the ionospheric F$_1$ and F$_2$ regions have different solar
  activity dependencies in terms of the sunspot group (SG) numbers: F$_1$ region critical
  frequency ($f_0F_1$) peaks at the same time with the small SG numbers, while the $f_0F_2$
  reaches its maximum at the same time with the large SG numbers, especially during the solar
  cycle 23. The observed differences in the sensitivity of ionospheric critical frequencies to
  sunspot group (SG) numbers provide a new insight into the solar activity effects on the
  ionosphere and space weather. While the F$_1$ layer is influenced by the slow solar wind,
    which is largely associated with small SGs, the ionospheric F$_2$ layer is more sensitive
    to Coronal Mass Ejections (CMEs) and fast solar winds, which are mainly produced by large
    SGs and coronal holes. The SG numbers maximize during of peak of the solar cycle and the
    number of coronal holes peaks during the sunspot declining phase. During solar minimum
  there are relatively less large SGs, hence reduced CME and flare activity. These results
  provide a new perspective for assessing how the different regions of the ionosphere respond
  to space weather effects.
\end{abstract}

\section{\textbf{Introduction}}
\label{sec:intro}
Sunspots are dark, cold and magnetically dense structures observed on the solar photosphere,
the visible surface of Sun. They have been observed systematically since 1610
\citep{Vaquero07, Clette_etal14}. Generally, they are observed on the solar disc as groups and
these groups have been classified according to their morphology, complexity, and evolution for
about a century \citep{Cortie1901, McIntosh90}. Using the number of observed groups and of
individual spots the daily International sunspot number (or Z\"urich number) is calculated by
\citep{Wolf1861}:
	\begin{equation}
	R_z = k (10g +f ),
	\label{eq:1}
	\end{equation}
where $f$ is the number of individual spots, $g$ is the number of observed sunspot groups, and
$k$ is a correction factor for each observatory. The sunspot number is the best known and the
longest solar activity index and is a good proxy for the solar activity variations
\citep{Clette_etal14, Hathaway15}.  These variations can also be represented by other solar
activity indicators, such as, sunspot areas (SSAs), sunspot group (SG) numbers, total solar
irradiance (TSI), and 10.7 cm solar radio flux (F$_{10.7}$).  The solar activity variations
obtained from these indicators show small differences in time, depending on the description of
indices and also the background physical mechanisms. Hence we will limit ourselves to the
analysis in time to SGs as the other indexes will follow more or less the same behavior.

R${_z}$ suggests that the daily sunspot number is directly related to the observed group and
individual sunspot numbers, without taking into account group/sunspot properties. However, a
robust result presented in the work by \citet{Kilcik_etal11a} indicates that the number of
large SGs peaks about two years later than the small ones. It has also been found that large
group numbers show a better correlation with the maximum speed of coronal mass ejections
(CMEs) and geomagnetic indices ($Ap$ and $Dst$) than the small ones for the solar cycle 23
\citep{Kilcik_etal11b}.

Solar activity variations strongly affect Earth's magnetosphere, ionosphere, and
thermosphere. The terrestrial thermosphere-ionosphere system is extremely variable due
primarily to the influence of lower atmospheric internal waves from below \citep[][and
references therein]{YigitMedvedev15} and geomagnetic and solar activity variations from above
\citep[e.g,][]{SmithtroSojka05b, Yigit_etal16}. Also, solar cycle variations may influence the
propagation and dissipation of gravity waves in the thermosphere
\citep{YigitMedvedev10}. 

Extending from $\sim$50 to $\sim$1000 km, the ionosphere forms the partially ionized portion
of the neutral upper atmosphere. A number of communication and navigation systems use the
ionosphere in which radio signals are propagated and transmitted. Each ionospheric layer has a
maximum frequency, known as critical frequency, at which radio waves can be transmitted
through and be reflected back (for normal incidence) to Earth most efficiently. The ionosphere
is transparent to the radio waves at frequencies higher than the critical frequency while
waves will be reflected back to Earth at frequencies lower than the critical
frequency \citep{Elias_etal17}.

\begin{figure*}[t]
	\centering
	\includegraphics[width=0.6\textwidth,angle=-90]{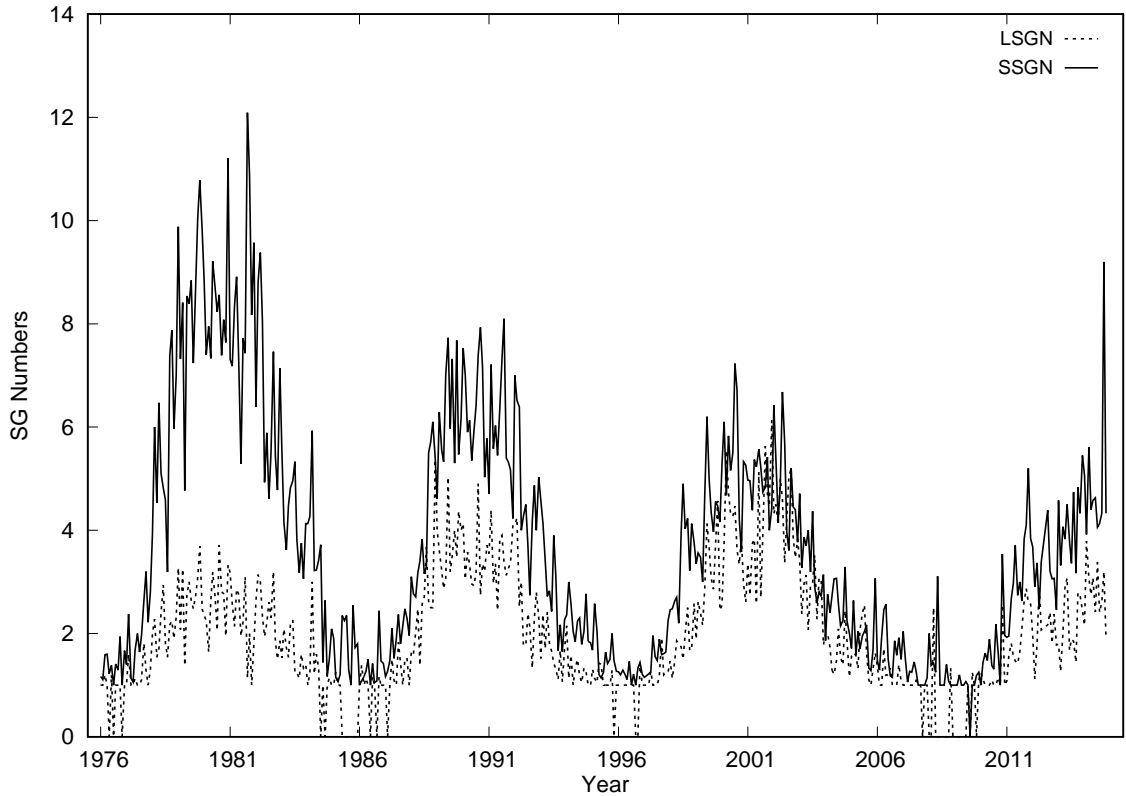}
	\caption{Monthly average large sunspot group number (LSGN) and small sunspot group number (SSGN) data for the last four solar cycles.}
	\label{fig:figssn}
\end{figure*}

\begin{figure}[t]
	\centering
	\includegraphics[width=0.7\columnwidth,angle=-90]{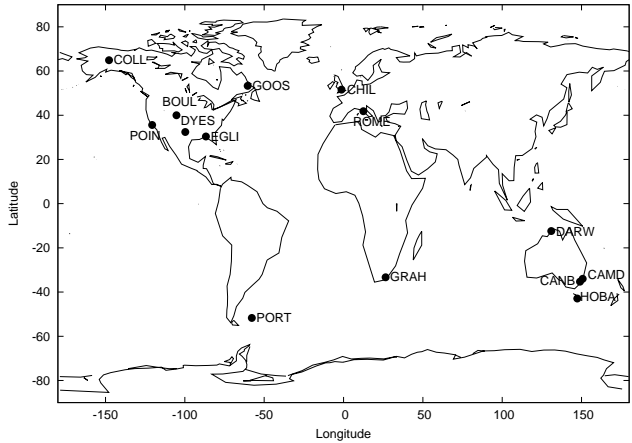}
	\caption{Global distribution of the ionospheric stations where the ionospheric
          critical frequencies (f$_o$F$_1$, and f$_o$F$_2$) are observed.}
	\label{fig:fig1}
\end{figure}

\begin{figure*}[t]
	\centering
	\includegraphics[width=0.7\textwidth,angle=-90]{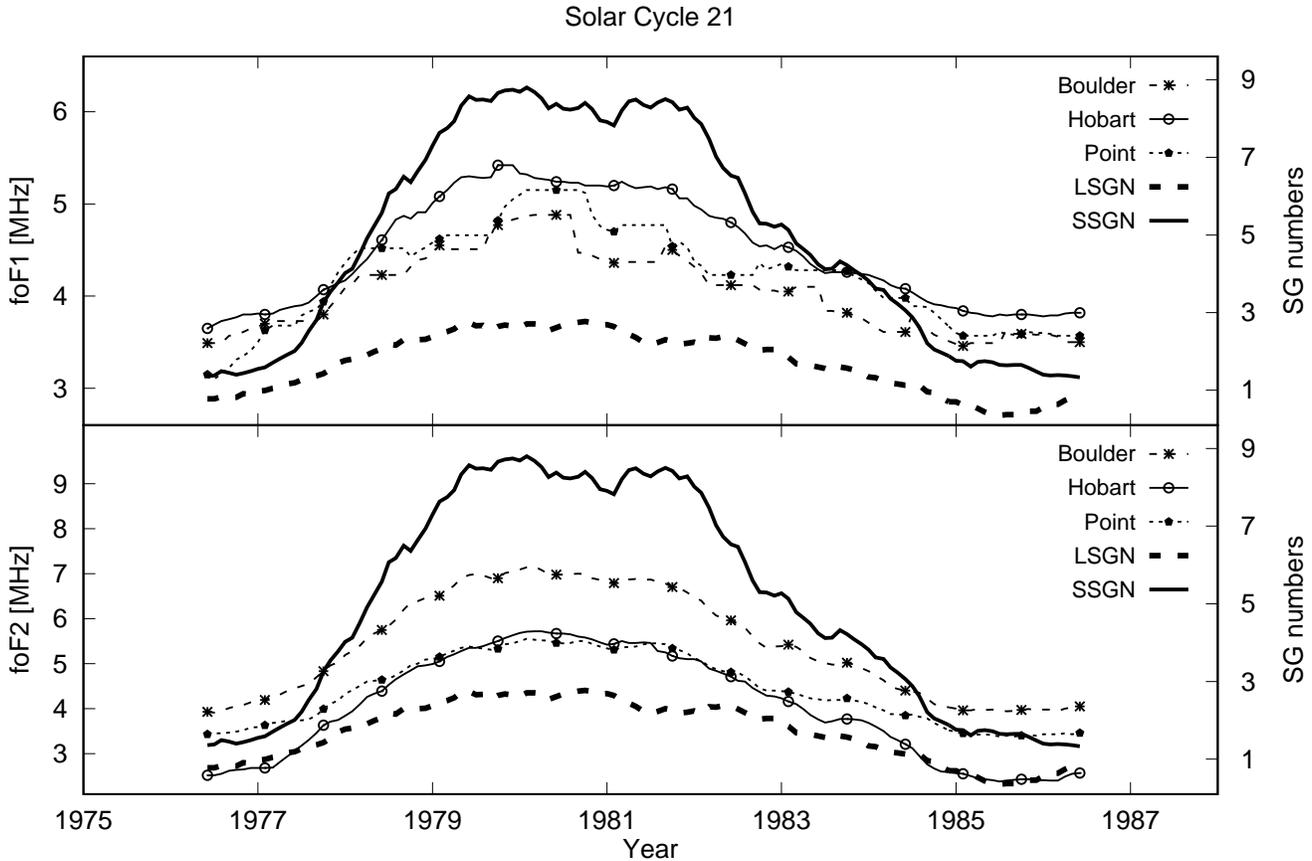}
        \caption{Temporal variations of the large (dashed line, LSSN) and small (solid line,
          SSSN) sunspot group numbers along with the variations of the ionospheric critical
          frequencies $f_0F_1 $ (top panel) and $f_0F_2 $ (bottom panel) in units of MHz
          during solar cycle 21. The different ionospheric stations are overplotted with
          different lines. }
	\label{fig:fig2}
\end{figure*}

\begin{figure*}[t]
	\centering
	\includegraphics[width=0.7\textwidth,angle=-90]{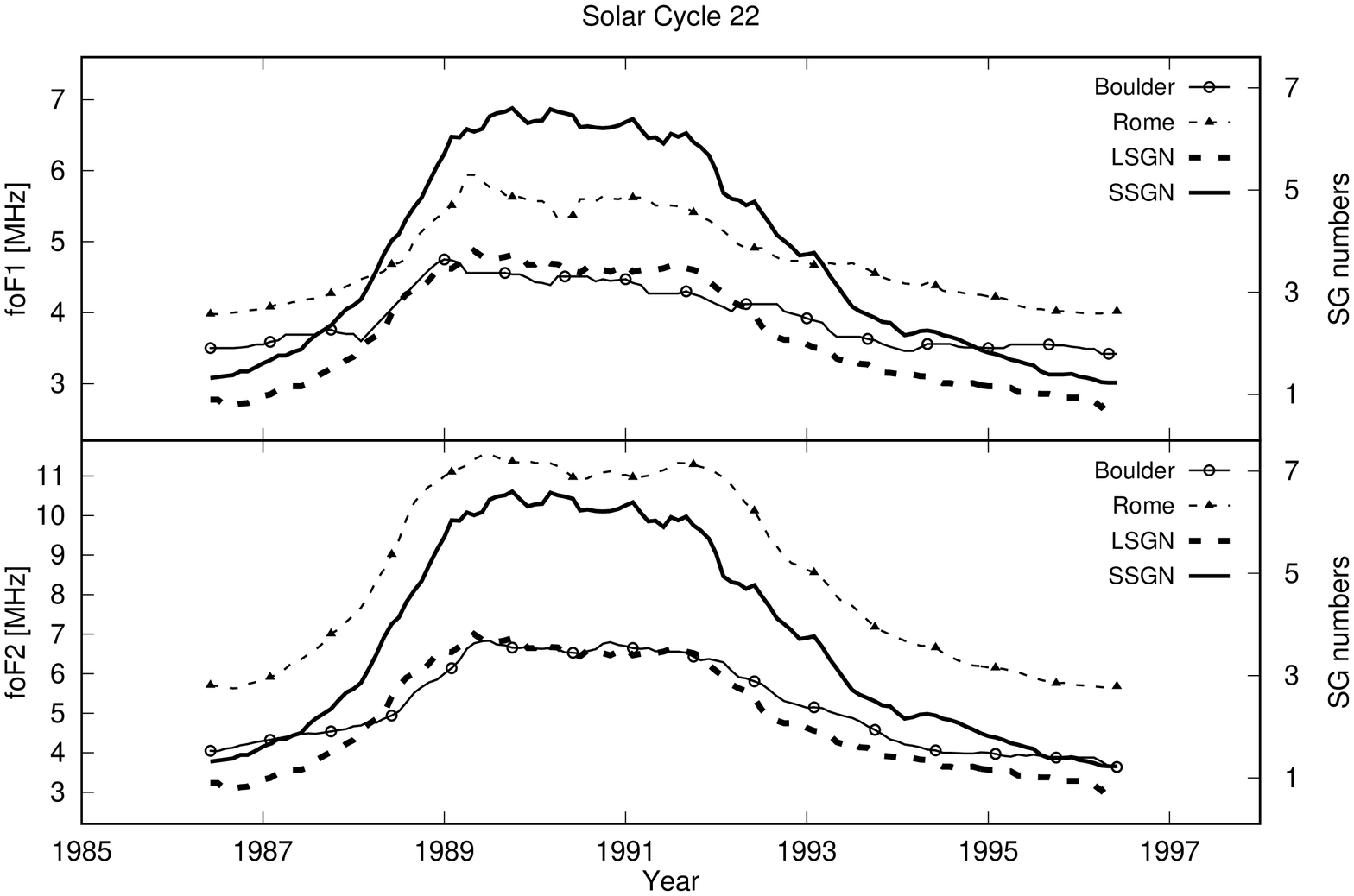}
	\caption{Same as Figure \ref{fig:fig2} but solar cycle 22.}
	\label{fig:fig3}
\end{figure*}

\begin{figure*}[t]
	\centering
	\includegraphics[width=0.7\textwidth,angle=-90]{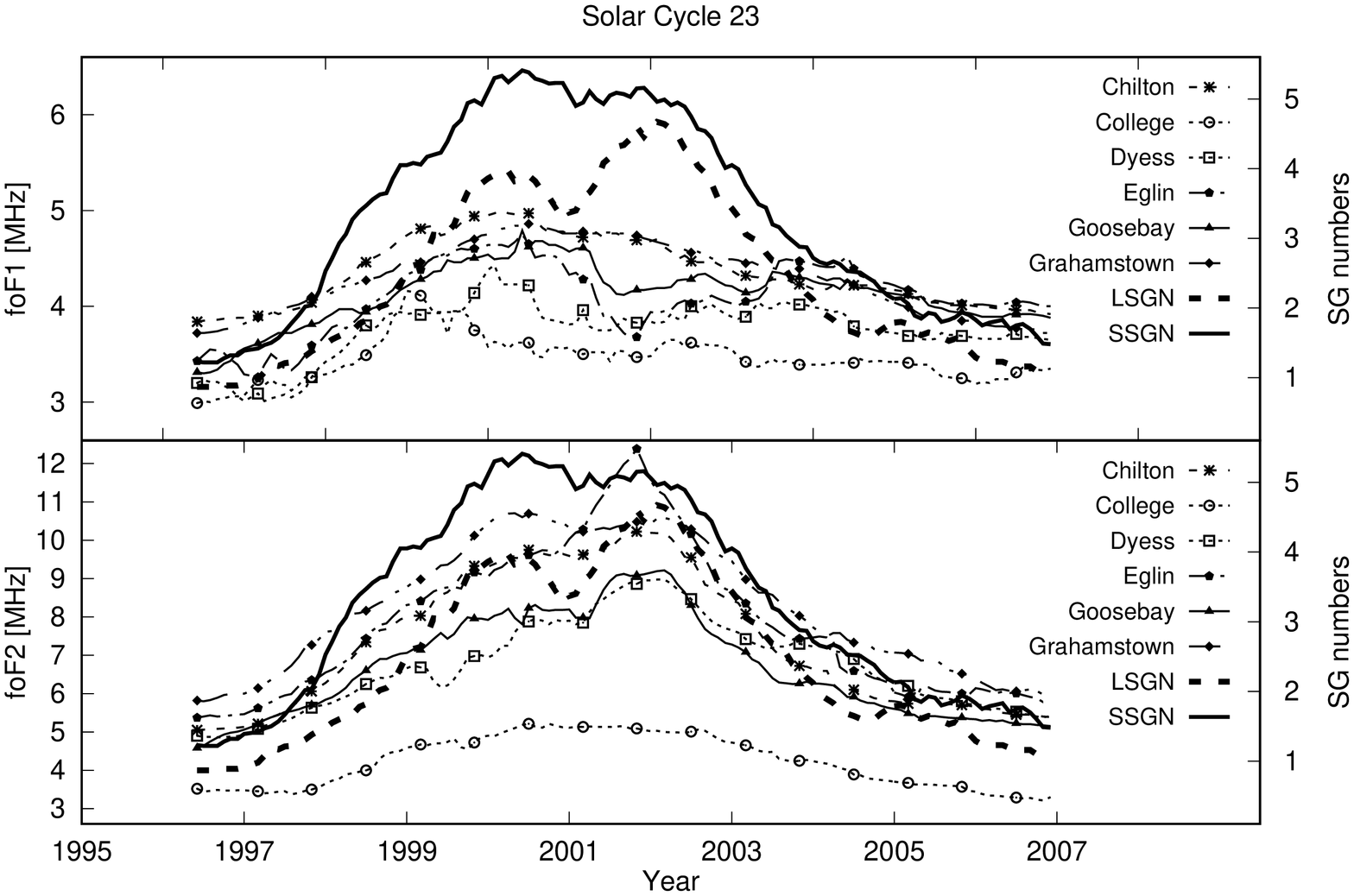}
	\caption{Same as Figure \ref{fig:fig2} but solar cycle 23.}
	\label{fig:fig4}
\end{figure*}

\begin{figure*}[t]
	\centering
	\includegraphics[width=0.7\textwidth,angle=-90]{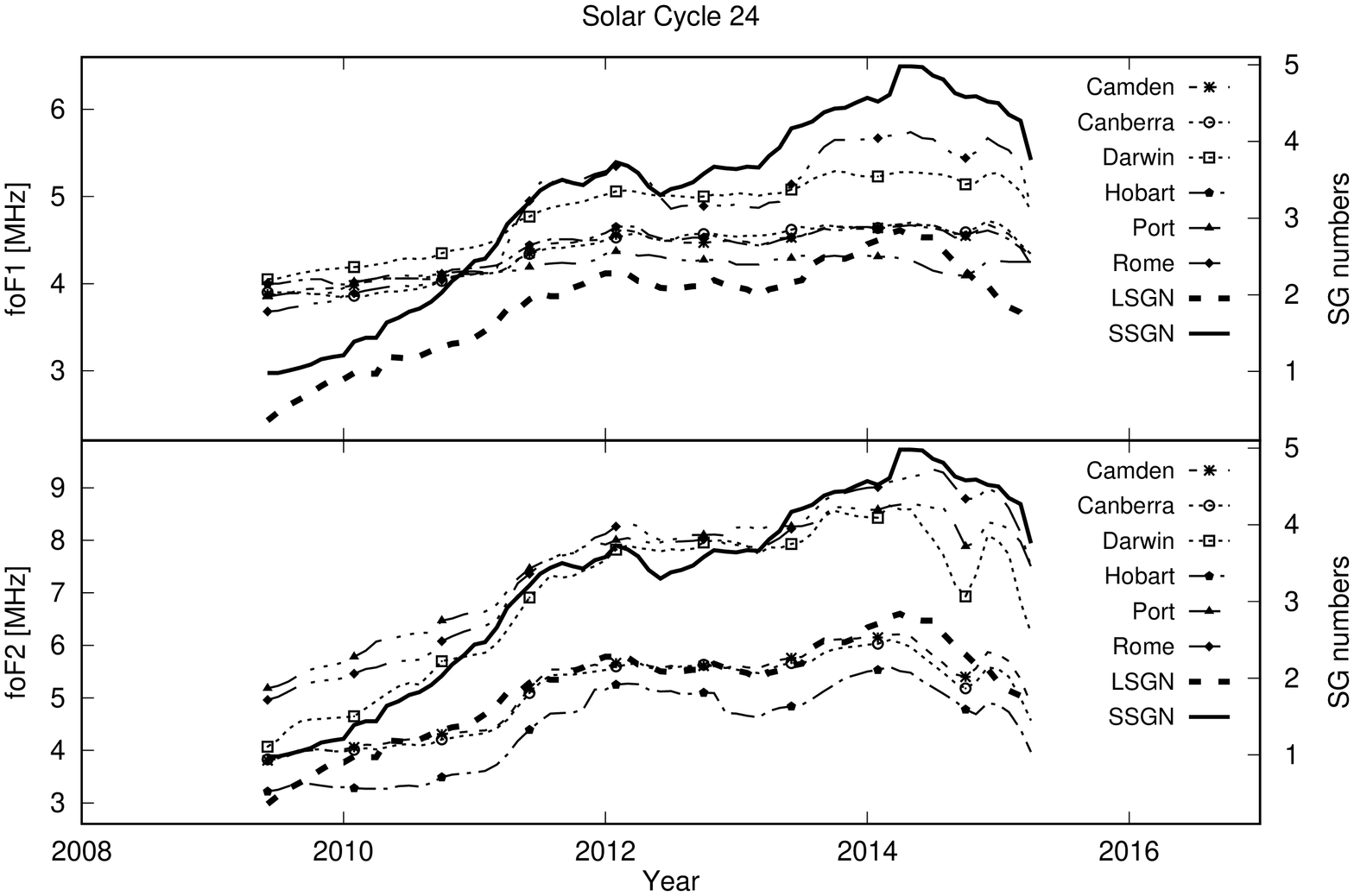}
	\caption{Same as Figure \ref{fig:fig2} but solar cycle 24.}
	\label{fig:fig5}
\end{figure*}

\begin{table*}[]
  \centering
  \caption{Correlation analysis results for each station and solar data set for
  the last four solar cycles. Errors are calculated by using Fisher test.}
  \label{tb1}
  \begin{tabular}{c|c|c|c|c}
    \hline
    & \multicolumn{2}{c|}{\bf{foF1}} & \multicolumn{2}{c}{\bf{foF2}}                    \\ \hline
    Station     & LSGN            & SSGN            & LSGN             & SSGN           \\ \hline
    \multicolumn{5}{c}{\it{Solar Cycle 21}}                                             \\ \hline
    Boulder     & 0.94 $\pm$ 0.02 & 0.94 $\pm$ 0.02 & 0.97 $\pm$ 0.01 & 0.99 $<$ 0.01   \\
    Hobart      & 0.96 $\pm$ 0.02 & 0.99 $<$ 0.01   & 0.98 $\pm$ 0.01 & 0.99 $<$ 0.01   \\
    Point       & 0.98 $\pm$ 0.01 & 0.94 $\pm$ 0.02 & 0.97 $\pm$ 0.01 & 0.99 $<$ 0.01   \\ \hline
    \multicolumn{5}{c}{\it{Solar Cycle 22}}                                             \\ \hline 
    Boulder     & 0.97 $\pm$ 0.01 & 0.96 $\pm$ 0.02 & 0.97 $\pm$ 0.01 & 0.99 $<$ 0.01   \\
    Rome        & 0.98 $\pm$ 0.01 & 0.98 $\pm$ 0.01 & 0.99 $<$ 0.01   & 0.99 $<$ 0.01   \\ \hline
    \multicolumn{5}{c} {\it{Solar Cycle 23}}                                            \\ \hline
    Chilton     & 0.86 $\pm$ 0.05 & 0.96 $\pm$ 0.02 & 0.98 $\pm$ 0.01 & 0.99 $<$ 0.01   \\
    College     & 0.61 $\pm$ 0.12 & 0.72 $\pm$ 0.10 & 0.95 $\pm$ 0.02 & 0.97 $\pm$ 0.01 \\
    Dyess       & 0.68 $\pm$ 0.11 & 0.71 $\pm$ 0.10 & 0.92 $\pm$ 0.03 & 0.86 $\pm$ 0.05 \\
    Eglin       & 0.44 $\pm$ 0.15 & 0.53 $\pm$ 0.14 & 0.96 $\pm$ 0.02 & 0.94 $\pm$ 0.02 \\
    Goosebay    & 0.75 $\pm$ 0.09 & 0.82 $\pm$ 0.07 & 0.97 $\pm$ 0.01 & 0.97 $\pm$ 0.01 \\
    Grahamstown & 0.93 $\pm$ 0.03 & 0.98 $\pm$ 0.01 & 0.97 $\pm$ 0.01 & 0.99 $<$ 0.01   \\ \hline
    \multicolumn{5}{c}{\it{Solar Cycle 24}}                                             \\ \hline
    Camden      & 0.96 $\pm$ 0.02 & 0.97 $\pm$ 0.02 & 0.95 $\pm$ 0.03 & 0.95 $\pm$ 0.03 \\
    Canberra    & 0.94 $\pm$ 0.04 & 0.97 $\pm$ 0.02 & 0.95 $\pm$ 0.03 & 0.92 $\pm$ 0.05 \\
    Darwin      & 0.96 $\pm$ 0.02 & 0.98 $\pm$ 0.01 & 0.96 $\pm$ 0.02 & 0.94 $\pm$ 0.03 \\
    Hobart      & 0.96 $\pm$ 0.02 & 0.94 $\pm$ 0.03 & 0.95 $\pm$ 0.03 & 0.91 $\pm$ 0.05 \\
    Port        & 0.85 $\pm$ 0.08 & 0.80 $\pm$ 0.10 & 0.97 $\pm$ 0.02 & 0.97 $\pm$ 0.02 \\
    Rome        & 0.95 $\pm$ 0.03 & 0.97 $\pm$ 0.02 & 0.96 $\pm$ 0.02 & 0.98 $\pm$ 0.01 \\ \hline
  \end{tabular}
\end{table*}

Photoionization, which is responsible for the formation ion-electron pairs, is dependent on
  the presence of the type of ionizable species, which can be influenced by other radiative
  loss process. The F$_1$ layer is primarily photochemically controlled. \ce{O+} ions dominate
  as a consequence of the photoionization of neutral atomic oxygen, while it is lost by
  ion-molecule interchange with \ce{O2} and \ce{N2}.  The F$_2$ region is rather dynamically
  controlled as a transition from chemical control to diffusive transport takes place. These
  dynamical processes include ambipolar diffusion, wind-induced drifts along the geomagnetic
  field, and wave effects. The peak ion content in the F-region occurs at the location where
  chemical and diffusive effects are of equal importance, which leads to the formation of the
  F$_2$ layer.

Overall, the F$_1$ layer of the ionosphere exhibits dependence on the solar zenith angle,
season, and geomagnetic activity. Therefore, it is more pronounced in the summer than in the
winter, disappears during the night and sometimes during the winter days. On the contrary, the
F$_2$ layer, where globally the largest amount of plasma is found, is a permanent feature of
the ionosphere under all solar-terrestrial conditions. 

Due to the importance of ionospheric critical frequencies for communication and their
relationship with solar activity, it has been studied extensively by a number of authors
\citep[e.g.,][and references therein]{Forbes_etal00, Kane06, Chakrabarty_etal14}. Most of these
studies compared the foF$_2$ critical frequency with solar activity variations in various time
scales.

In this study, in order to better understand the response of the different ionospheric regions to
solar variations, we examine the long-term temporal dependencies of ionospheric $f_0F_1$ and $f_0F_2$
critical frequencies to the SG numbers during the last three complete solar cycles (cycles 21, 22, and
23) and the ascending and maximum phases of solar cycle 24. Specifically, the temporal variations of
$f_0F_1$ and $f_0F_2$ are compared with the variations of small and large sunspot groups. It is
assumed that the connection is not due to the sunspots in themselves, but through the fluctuations of
the activity that way generated. Thus we have emphasized this aspect, and in particular, by
considering the double peaks which sometimes occurred in the solar activity not explained so far, that
may affect the behavior or corpuscular emissions (solar wind, CMEs, etc.) as well as the radiative
effects.

The structure of the paper is as follows: Next Section (Section \ref{sec:data}) briefly describes the
data to be used in our analysis; Section \ref{sec:results} presents the results on the relationship
between the ionospheric critical frequencies and the sunspot groups, and Section \ref{sec:conc}
presents the conclusions and briefly discusses their implications.

\section{\textbf{Data}}\label{sec:data}
\subsection{The Sunspot Group (SG) Number Data} 
The SG number data were downloaded from the National Geophysical Data Center
(NGDC)\footnote{http://www.ngdc.noaa.gov/stp/space-weather/solar-data/solar-features/sunspot-regions/}
for each recorded group during the observed day. The data were collected by the United States
Air Force/Mount Wilson Observatory (USAF/MWL) since 1982 and the previous data are taken from
Rome and Taipei Observatories. The USAF/MWL database also includes measurements from the
Learmonth Solar Observatory, the Holloman Solar Observatory, and the San Vito Solar
Observatory. We used the Learmonth station data as the principal data source for the last three
cycles (since 1986), and gaps were filled with records from one of the other stations listed
above. Thus, a nearly continuous daily SG number data set was produced, according to the
Modified Zurich Sunspot Classification, for both large (D,E,F) and small (A, B, C, H)
groups. Due to the time coverage of USAF/MWL data set, the Rome observatory data are used as a
reference data for cycle 21 and gaps in this data set were filled with Taipei observatory
data. But, this data set still has many gaps compared to Learmonth data.  As a final step, the
monthly mean values for the large and small SG numbers were calculated. Temporal variation of the monthly large and small SGs is presented in Figure \ref{fig:figssn}. 
To remove the short term fluctuations and reveal the long term trend 12-step running average smoothing was applied and used in the
analysis.

\subsection{Ionospheric Critical Frequencies Data}
The ionospheric critical frequencies data for the selected stations are taken from Space
Physics Interactive Data Resource, SPIDR\footnote{http://spidr.ionosonde.net/spidr/}. To
  select these stations the main criteria was existence of continuous data for the investigated
  solar cycles. We used smoothed monthly median critical frequencies, $f_0F_1$ and $f_0F_2$,
recorded at 14 stations distributed over the globe (see Figure \ref{fig:fig1}) within the
period of 1976--2015, which cover three full solar cycles (cycle 21, 22, and 23) and ascending
and maximum phases of the solar cycle 24. Due to the lack of continuous station data our
  analysis could not cover all latitude intervals. Monthly median values calculated from data
taken at 14:00 LT for each day of a given month for the investigated time period.  In general,
the $f_0F_2$ data have much better temporal coverage, while the number of total observing days
strongly decreased during the winter times for the $f_0F_1$ data, as expected.  But a few days
of observation still exist. To calculate the monthly median values we used all existing data
for each month. To remove the short term fluctuations due to gaps (especially in $f_0F_1$) in
monthly median data and reveal the long term trend we used 12-step running average smoothing
method. Three stations for cycle 21, two stations for cycle 22 and six stations for cycle 23
and 24 were analyzed because of the lack of continuous data.

\section{\textbf{Analysis and Results}}
\label{sec:results}

In this study, first, we compared temporal variation of SG numbers and ionospheric critical frequencies. Second we calculated Pearson correlation coefficients and their confidence levels between these data sets. For the confidence level Fisher's test which gives upper and lower bounds of correlation coefficients, were used. We used the highest error as an error level.

Figure \ref{fig:fig2} presents the temporal variations of the observed $f_0F_1$ (top panel),
$f_0F_2$ (bottom panel), small sunspot group numbers (SSGNs), and large sunspot group numbers
(LSGNs) during the solar cycle 21. Thick solid and dashed lines denote the SSGNs and LSGNs,
respectively. The other lines represent the different stations. Note that the SSGN and LSGN
variations are overplotted in both the top and bottom panels to facilitate a better comparison
with the critical frequencies. The subsequent Figures \ref{fig:fig3}, \ref{fig:fig4}, and
\ref{fig:fig5} present the associated temporal variations for the solar cycles 22, 23, and 24,
respectively, in a manner similar to Figure \ref{fig:fig2}. Overall, we have a coverage
approximately from 1976 to 2015. Analysis of the different solar cycle behavior of the both
critical frequencies and the SG numbers suggest that SSGN and LSGN data show almost
similar variations during the solar cycles except for solar cycle 23.  The LSGN peaks about one
to two years later than the SSGN during solar cycle 23, while both large and small SG numbers
peak at almost the same time or the differences between two maxima are not such a prominent
feature during solar cycles 21, 22, and 24.  In general, $f_0F_1$ and $f_0F_2$
demonstrate different solar cycle variations compared with the variations of the SGs, which
suggests that the different ionospheric regions are sensitive to the activity of the different
regions in the solar photosphere.

Overall, in terms of temporal variations, the $f_0F_1$ follows the small SG numbers, while the
$f_0F_2$ follows the large SG numbers. The marked difference between large and small SGs
temporal behavior during a cycle is clearly seen during solar cycle 23 (Figure \ref{fig:fig4}),
during which the occurrence of the strong SSGN global maximum coincides with the LSGN local
maximum in 2000 and the occurrence of the weak local maximum coincides with the LSGN global
maximum in 2002, demonstrating the $\sim$2-year preceding of the peak SSGN with respect to the
LSGN.  In general, intercomparison of the critical frequency trends with the SG
numbers suggest that the $f_0F_2$ data follows the large SG number variations, while the
variations of $f_0F_1$ rather follow the small SG numbers. These findings are pronounced much
more clearly in the solar cycle 23 (see Figures 2, 3, 4, and 5). However, during the other
solar cycles, this effect is masked out because the LSGN and SSGN have similar temporal
variations (there are no prominent differences between the two peaks) in terms of the timing of
their global maxima. Overall, these results suggest that the ionospheric $f_0F_1$ and $f_0F_2$
critical frequencies respond differently to the different origin of the solar activity,
characterized by the different sunspot groups. Essentially, most of the active events such as
solar flares and CMEs occur in the large/complex active regions, mainly populated by large
sunspots \citep{Kilcik_etal11b}. Thus our results, in particular related to the cycle 23,
indicate that $f_0F_2$, and thus the $F_2$ region, is sensitive to active sun
(flares, CMEs), while the $f_0F_1$, and thus the $F_1$ region, is sensitive to quite sun
(regular solar wind).

In Table \ref{tb1} we presented correlation coefficients between 12-step running averaged
  SG numbers and ionospheric critical frequencies for each cycle, separately. As shown in this
  table, generally correlation coefficients between $f_0F_1$ and $f_0F_2$ critical frequencies
  and large and small SG numbers are comparable during all cycles except cycle
  23. During solar cycle 23 large groups are well correlated with $f_0F_2$, while small groups
  with $f_0F_1$ for most of the stations.

\section{\textbf{Discussion and Conclusions}}
\label{sec:conc}

The long-term temporal variations of the observed ionospheric $f_0F_1$ and $f_0F_2$ critical
frequencies have been investigated for the last four solar cycles from the cycles 21 to 24
(1976 through 2015) and been compared with the associated variations of the solar activity
represented by small (SSGNs) and large (LSGNs) sunspot group numbers. The same analysis may
well have been carried out by other solar activity indicators such as TSI, $F_{10.7}$ etc.,
but it is known that their temporal variations were highly correlated by the sunspot
number. We preferably have sought to place the emphasis on the importance of SG numbers due to
their separability into small and large structures and thus their insightful physical
association with coronal mass ejections (CMEs) and solar flare activity
\citep{Kilcik_etal17}. Our key finding is that the temporal variations of the ionospheric
critical frequencies exhibit different solar cycle variations in terms of the SSGN and LSGN
for the investigated time periods, in particular during cycle 23 (1996 through 2008).

Close investigation of the solar cycle 23 demonstrates an anomalous character compared to the
solar cycles 21, 22, and 24. Namely, the global maximum of the SSGN occurs about 2 years
earlier than the global maximum of LSGN. Interestingly, the $f_0F_1$ peak occurs at the same
time as the occurrence of the global maximum of SSGN, while the $f_0F_2$ peak occurs at the
same as the global maximum of LSGN. This effects is not depictable in the other cycles as the
SSGN and LSGN generally maximize at the same time.

The possible long-term drivers of F$_1$ and F$_2$ regions critical frequencies include
long-term variations in solar and geomagnetic activity, various greenhouse gases (e.g., CO$_2$,
CH$_4$) concentrations, ozone variations, water vapor and the magnetic field variations
\citep[][and references therein]{ Yue_etal06, Mikhailov08, Lastovicka_etal12,
  Gordiyenko_etal14}. Here we focused exclusively on the solar effects on the
  ionospheric critical frequencies.

There has been previous scientific evidence that the solar cycle 23 was indeed an anomalous
cycle. For instance, \citet{deToma_etal04} found that the magnitude of TSI during the solar
cycle 23 was comparable to solar cycle 22, while the magnitude of the ISSN was much
lower. \citet{Kilcik_etal11a} found that the facular area was also lower during the solar
cycle 23. Contrary to small groups, which were strongly diminished during solar cycle 23, the
number of large groups were comparable to, or even higher, than that of solar cycle 22
\citep{LefevreClette11, Kilcik_etal11a, Kilcik_etal14}.  Also many low-latitude coronal holes
observed during the declining phase of cycle 23 \citep[see][] {Abramenko_etal10}. It can be also
noted that \citet{Gordiyenko_etal14} found similar results for the annual means of $f_0F_2$
(see Figure 4 in their paper). On the other hand, it is known that all solar activity
indicators such as the TSI, SSA, F10.7, etc., peaked in 2002 during solar cycle 23. Also,
similar to the ISSN, they all show double/multiple peaks near the maximum of this cycle.

\citet{Kilcik_etal11a} investigated SG numbers in two categories, as large and small, from
1964 to 2008. They found that the number of large groups peaked about two years later than
small ones except for solar cycle 22 (1986--1996): the difference between large and small SG
numbers is very prominent during solar cycle 23, while maxima were almost flat during solar
cycle 21 and 22.  Recently, \citet{Kilcik_etal14} analyzed the sunspot counts (SSCs) in four
categories, as small, medium, large and final, from 1982 to 2014, and found similar results
for SSCs. Here, we analyzed monthly median $f_0F_1$ and $f_0F_2$ for 1976 - 2015 time interval
which include this anomalous solar cycle 23 (1996--2008) and which clearly reveals that during
solar cycle 23 the $f_0F_1$ is following the temporal variation of SSGN, while the $f_0F_2$ is
following the LSGN. Both $f_0F_1$ and $f_0F_2$ have flat peaks during solar cycles 21 and 22 and
they peaked in the second maximum of solar cycle 24 similar to SG numbers. Thus, our results
for the anomalous solar cycle 23 along with its comparison with the cycles 21, 22, and 24
provide further insight into the effects of solar activity on the ionosphere. Specifically,
they indicate that variations of the two different sunspot categories have different effect on
these ionospheric layers: the first one is the geo-effective solar events (flares, CMEs, etc.)
which are mainly produced by large/complex sunspot groups \citep{Eren_etal17}. They mostly
describe the active sun and are more effective on the ionospheric F$_2$ layer critical
frequency. The second one relating to the small SGs may produce quite rare flares compared to
the large ones \citep{Lee_etal12}.  These sunspot groups may describe the quite/weak solar
activity.

The ionospheric F$_1$ and F$_2$ layers have different physical characteristics. F$_1$ layer
is photochemically controlled and  behaves like a Chapman layer, demonstrating
overall a $\cos\chi $ (solar zenith angle) dependence, while the F$_2$ layer can substantially departs from a simple
solar control due to the relative significance of dynamical processes. Also, the
ionospheric F$_2$ layer is very susceptible to geomagnetic activity and CMEs
\citep{Burns_etal07}. It is the most anomalous and variable, hence the least predictable
ionospheric layer due to the complex interplay of chemistry, dynamics (e.g., diffusive
transport), and coupling to electric fields of magnetospheric origin. Generally, the
characteristics and intensity of CMEs and solar winds can be associated with the scales of the
SGs, providing a link between the ionospheric F-region and the solar atmosphere. Thus,
especially our analysis pertaining to cycle 23 lead to the conclusion that the long-term variations of the ionospheric F$_2$ layer is influenced more by
the variations of LSGNs, while the long-term variations of F$_1$ layer is linked more to the
variations of the SSGNs. In other words, this implies that ionospheric $F_2$ layer is more
sensitive to CMEs and fast solar winds (active sun conditions), while the $F_1$ layer is more
sensitive to slow/regular solar wind (quite sun conditions).

\acknowledgements{
The sunspot group data were taken from the National Geophysical Data Center (NGDC) web
page. The f$_0$F$_1$ and f$_0$F$_2$ data sets were retrieved from the Space Physics
Interactive Data Resource (SPIDR) web page. This study was supported by the Scientific and
Technical Council of Turkey by the Project of 115F031. One of us (JPR) acknowledges the
International Space Science Institute (ISSI) in Bern (Switzerland) for a ``visitor scientist"
grant. EY was partially funded by the NSF grant AGS 1452137.}
  
%
%

\end{document}